\documentclass[sigplan, nonacm]{acmart}
\usepackage{underscore}

\usepackage{algorithm}
\usepackage{algpseudocode}
\usepackage{amsmath}
\usepackage{makecell}
\usepackage{subcaption}
\usepackage{tikz}
\usepackage{xcolor}
\definecolor{codebg}{HTML}{F7F7F7}
\definecolor{commhop}{RGB}{36,140,64}
\definecolor{commtree}{RGB}{20,150,165}
\definecolor{commfact}{RGB}{210,140,40}
\newsavebox{\codepanelbox}
\newenvironment{codepanel}{%
  \begin{lrbox}{\codepanelbox}%
  \begin{minipage}{\dimexpr\linewidth-8pt\relax}%
}{%
  \end{minipage}%
  \end{lrbox}%
  \setlength{\fboxsep}{4pt}%
  \colorbox{codebg}{\usebox{\codepanelbox}}%
}
\usetikzlibrary{arrows.meta, positioning, calc}
\AtBeginDocument{%
  }

\begin{document}

\title{Cyclotron: A Recurrence Language for Expressing Interprocessor Communication}

 \author{Shiv Sundram}
 \email{shiv1@stanford.edu}
 \affiliation{%
   \institution{Stanford University}
   \city{Stanford}
   \state{California}
   \country{USA}
 }

 \author{Akhilesh Balasingam}
 \email{avb03@stanford.edu}
 \affiliation{%
   \institution{Stanford University}
   \city{Stanford}
   \state{California}
   \country{USA}
 }

 \author{Nathan Zhang}
 \email{stanfurd@stanford.edu}
 \affiliation{%
   \institution{Stanford University}
   \city{Stanford}
   \state{California}
   \country{USA}
 }

 \author{Kunle Olukotun}
 \email{kunle@stanford.edu}
 \affiliation{%
   \institution{Stanford University}
   \city{Stanford}
   \state{California}
   \country{USA}
 }

 \author{Fredrik Kjolstad}
 \email{kjolstad@stanford.edu}
 \affiliation{%
   \institution{Stanford University}
   \city{Stanford}
   \state{California}
   \country{USA}
 }


\renewcommand{\shortauthors}{Anonymous}

\begin{abstract}
We present Cyclotron, a compiler that expresses streaming dataflow algorithms
using recurrence equations, which are then compiled to distributed topologies
of interlinked processors. Our compiler lowers a language of recurrences over logical
iteration spaces into per-processor programs of sends, receives,
and computations. In these recurrences, data accesses are not limited to external accesses to memory; data can live separately on the iteration space itself, allowing internal points to  communicate values directly with other points. Users specify this communication directly in the  recurrence language, which allows them to express various communication patterns such as  a nearest-neighbor systolic exchange, a one-to-all broadcast, a bidirectional split ring, or an asymmetric and piecewise broadcast pattern, which are often not supported by standard distributed linear algebra libraries. These recurrences describe the algorithm and communication independently of how they map onto the hardware. A programmer then chooses which dimensions execute across processors and which execute over time, without rewriting the entire program. Users only specify data dependencies across the iteration space, and the compiler safely generates a program, per process, with correct placement of both communication and computational operators according to the chosen space--time mapping. We evaluate distributed matrix multiplication, TRSM, Cholesky, and fused Cholesky--TRSM on up to 16 CPU nodes and 16 GPUs, comparing against ScaLAPACK, SLATE, and cuSolverMp. Cyclotron is competitive with and often outperforms these hand-optimized libraries; for Cholesky, the best communication strategy changes with problem size and architecture.
\end{abstract}

\received{2 June 2026}

\maketitle

\section{Introduction}
\label{sec:intro}

Generations of architectural improvements have widened the gap between the cost
of data movement and the cost of computation. Performance optimization is now
dominated by minimizing memory traffic and network communication on distributed architectures. 
Extracting performance from such hardware is largely a matter of orchestrating how data
moves through that fabric: laterally between PEs, and temporally as values are
retained in fast memory across steps.

Yet the dominant abstractions for programming these machines do not represent data
movement beyond memory loads/stores. In tensor algebra notation, loop-based IRs, and polyhedral representations, every
data access is a tensor access---logical reads from or writes to memory
locations. Such a program says \emph{what} value each statement consumes, but
not \emph{how} or \emph{where} that value travels between processors or across
time. The movement is left implicit, to be recovered by dependence analysis and
then realized by a lowering pass.

In computational problems where communication is a primary bottleneck, such as in distributed linear algebra (e.g., matrix multiplication), tensor algebra, and recurrences, the best communication policy can vary with input size, tile size, interconnect, and processor architecture. Yet standard distributed libraries like ScaLAPACK \citep{choi1992scalapack} or SLATE \citep{gates2019slate} either hard-code communication policies or expose only a small set of predefined options. Consequently, the best communication schedule for a particular problem and architecture may not be available to the programmer.

Compiler frameworks avoid pitfalls of writing distributed code by hand, which is error-prone and can lead to indefinite hangs from mismatched or incorrectly ordered communication, while also being difficult to optimize (e.g., overlapping communication with computation).
Polyhedral frameworks have made important steps forward in modeling relationships between iteration spaces and memory accesses. Existing polyhedral frameworks, like AutoSA \citep{autosa} for systolic-array synthesis and AlphaZ for distributed-memory code generation,\cite{alphaz} lower dependencies into nearest-neighbor transfers; both therefore effectively commit to a systolic realization of communication. As a result, alternative realizations of the same logical dependence, such as a one-to-all broadcast, a tree-broadcast, or heterogeneous combinations of broadcast and forwarding, are not supported.

This limitation stems from what the user program leaves unspecified. With polyhedral frameworks, users write equations or loops over array accesses, and the compiler analyses the program to identify where communication should occur. There are typically many ways  to realize data movement, but without user input the compiler must choose among them. For example, AutoSA and AlphaZ only generate code where communication is limited to between neighboring PEs. This can leave performance on the table when a different communication strategy better matches the problem size or target architecture. Stellar \citep{genc2024stellar}, a framework for spatial accelerator design, allows users to specify dataflow and lets iteration-space points receive data from neighbors, supporting systolic patterns, but does not generalize beyond them.

This paper presents the Cyclotron compiler that makes the communication schedule
a first-class part of the input program. The program can express many different communication patterns as recurrence equations. These include systolic exchanges, one-to-all broadcast, tree-broadcast, bidirectional split rings, or asymmetric, piecewise combination of communication policies, in which different subdomains of a program and processor grid carry different communication policies. A user can specify that a data tile travel via nearest-neighbor hops in one part of the processor grid, via a tree broadcast in other parts, and bidirectionally in another.

\begin{figure}[t]
\centering
\includegraphics[width=\linewidth]{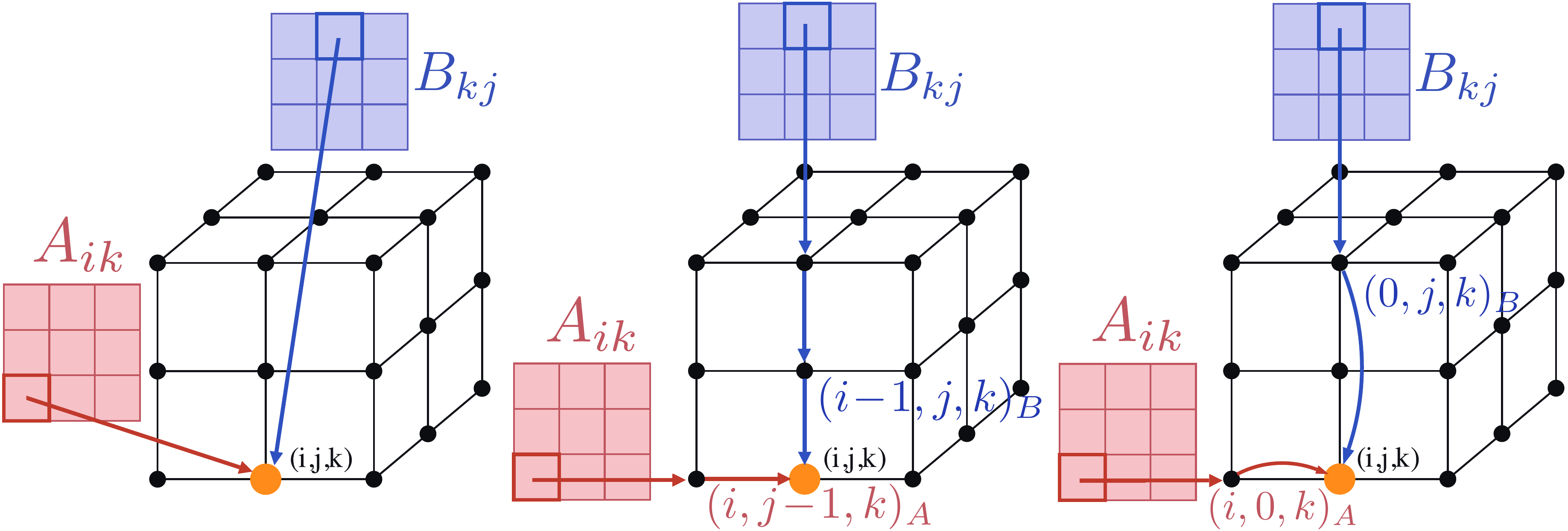}
\caption{Three spatial matrix multiplies. Left: each point reads operands
$A_{ik}$ and $B_{kj}$ directly from memory. Middle: each operand is loaded once
at a boundary face, and interior points receive $A$ and $B$ systolically
from neighbors. Right: each point receives its data directly from boundaries.
}
\label{fig:memory-vs-space-intro}
\end{figure}
In Cyclotron, the programmer writes
computation as a system of recurrences that can index into tensors that live on the iteration space in addition to tensors that live in memory, thus controlling data movement. Figure~\ref{fig:memory-vs-space-intro} illustrates the 3D iteration space of a matrix multiply, contrasting a program where data is accessed exclusively from tensors with two different programs where data is also obtained from points in the iteration space. What differs is how each value is obtained: directly from the tensor, from a neighboring iteration-space point, or from a boundary point via broadcasts. In Cyclotron, a value on the iteration space is named not only by which tensor element it is (e.g., $A_{ij}$) but by the point in the iteration space at which it lives, written
$(i,j,k)_{A}$. Movement, like the systolic transfer $(i,j,k)_A \leftarrow (i,j\!-\!1,k)_A$, is expressed in the same coordinate system as computation: a recurrence relating a value at one point to another point \emph{is} a data movement. A separate per-dimension annotation declares whether each dimension is spatial (movement between PEs) or temporal (retention across steps). Data movement is not limited to nearest neighbors, and because the recurrence is written independently of this annotation, the same program can be used to generate different spatial mappings.

The user's input recurrences are unordered. The ordering of compute and communication operations depends on which dimensions are mapped in space and time. Cyclotron's dependency algorithm uses this mapping to derive correct orderings, relieving users of the need to manually sequence such operations and avoiding indefinite waits caused by incorrect communication ordering.

These recurrences are compiled to distributed code through a fixed pipeline: a space--time mapping folds the iteration space and inserts a temporal loop. Our dependency-management algorithm resolves data dependencies into coupled sends and receives. Synchronization-aware control flow is inserted according to the target protocol. A final pass can apply a block-cyclic decomposition and deduplicate communication. 

Cyclotron lets programmers explore communication strategies with minimal program changes. The resulting code is competitive with, and often outperforms, expert-tuned libraries. We make the following contributions:
\begin{itemize}
  \item A recurrence-based programming model where communication is a first-class programmable concern, separate from computation and space--time mapping, enabling programmers to express and compose systolic, broadcast, ring, tree, and heterogeneous piecewise communication strategies.
  \item A dependency-management algorithm that uses the chosen space--time mapping to order computation and communication from unordered recurrences
  \item A complete compilation pipeline integrating the dependency algorithm with transformations that handle synchronization, block-cyclic decomposition and communication deduplication, generating efficient distributed programs.
\end{itemize}

By exploring different communication schedules, 
we demonstrate $1.7$--$1.8\times$ over ScaLAPACK for matrix multiply.
For Cholesky on up to 16 CPU nodes, we observe  $1.34$--$1.68\times$  speedup over ScaLAPACK  on CPUs
and $1.01-1.24 \times$ speedup over SLATE on GPUs (which uses CPU-side communication), and 1.12--1.28$\times$ speedup over cuSolverMp, Nvidia's optimized multi-GPU vendor library, when only using direct GPU-GPU communication. We demonstrate benefits of simultaneously composing different space--time mappings and how Cyclotron natively fuses kernels.

\section{Motivating Example}
\label{sec:motivation}
\label{sec:cholesky}

\begin{figure}[t]
\centering
\begin{minipage}[t]{0.49\linewidth}
\vspace*{0pt}
\begin{subfigure}[t]{\linewidth}
\centering
\scriptsize
\begin{codepanel}
\begin{algorithmic}
\For{$k = 0$ \textbf{to} $N{-}1$}
    \State $\text{POTRF}(L_{k,k})$
    \For{$i = k{+}1$ \textbf{to} $N{-}1$}
        \State $\text{TRSM}(L_{i,k}, L_{k,k})$
    \EndFor
    \For{$i = k{+}1$ \textbf{to} $N{-}1$}
        \State $\text{SYRK}(L_{i,i}, L_{i,k})$
        \For{$j = k{+}1$ \textbf{to} $i{-}1$}
            \State $\text{GEMM}(L_{i,j}, L_{i,k}, L_{j,k})$
        \EndFor
    \EndFor
\EndFor
\end{algorithmic}
\end{codepanel}
\caption{Right-looking $kij$.}
\label{fig:chol-kij}
\end{subfigure}

\vspace{0.6em}

\begin{subfigure}[t]{\linewidth}
\centering
\scriptsize
\begin{codepanel}
\begin{algorithmic}
\For{$i = 0$ \textbf{to} $N{-}1$}
    \For{$j = 0$ \textbf{to} $i{-}1$}
        \For{$k = 0$ \textbf{to} $j{-}1$}
            \State $\text{GEMM}(L_{i,j}, L_{i,k}, L_{j,k})$
        \EndFor
        \State $\text{SYRK}(L_{i,i}, L_{i,j})$
        \State $\text{TRSM}(L_{i,j}, L_{j,j})$
    \EndFor
    \State $\text{POTRF}(L_{i,i})$
\EndFor
\end{algorithmic}
\end{codepanel}
\caption{Up-looking $ijk$.}
\label{fig:chol-ijk}
\end{subfigure}
\end{minipage}\hfill
\begin{minipage}[t]{0.49\linewidth}
\vspace*{0pt}
\begin{subfigure}[t]{\linewidth}
\centering
\scriptsize
\begin{codepanel}
\begin{algorithmic}
\State $i,j \leftarrow \text{rank}()$
\For{$k = 0$ \textbf{to} $j$}
    \If{$k < j$}
        \State $\text{Recv } L_{i,k}\text{: }(i,\, j{-}1)$
        \State $\text{Send } L_{i,k}\text{: }(i,\, j{+}1)$
        \If{$i > j$}
            \State $\text{Recv } L_{j,k}\text{: }(i{-}1,\, j)$
            \State $\text{GEMM}(L_{i,j}, L_{i,k}, L_{j,k})$
        \Else
            \State $L_{j,k} \leftarrow L_{i,k}$
            \State $\text{SYRK}(L_{i,i}, L_{j,k})$
        \EndIf
        \State $\text{Send } L_{j,k}\text{: }(i{+}1,\, j)$
    \EndIf
    \If{$k = j,\ i > j$}
        \State $\text{Recv } L_{j,j}\text{: }(j,j)$
        \State $\text{TRSM}(L_{i,j}, L_{j,j})$
        \For{$d = 1$ \textbf{to} $N{-}j{-}1$}
            \State $\text{Send } L_{i,j}\text{: }(i,\, j{+}d)$
        \EndFor
    \EndIf
    \If{$k = j = i$}
        \State $\text{POTRF}(L_{i,i})$
        \For{$d = 1$ \textbf{to} $N{-}i{-}1$}
            \State $\text{Send } L_{i,i}\text{: }(i{+}d,\, i)$
        \EndFor
    \EndIf
\EndFor
\end{algorithmic}
\end{codepanel}
\caption{Per-processor lowering of $ijk$.}
\label{fig:chol-lowering}
\end{subfigure}
\end{minipage}

\caption{Two Cholesky loop nests (a,b) and a distributed lowering of the
up-looking nest (c). The order of the compute statements differs between the
two sequential variants, as do the number of inner loops and tensor indexing. The 
distributed nest (c) adds send/recv statements. $\text{Recv}\,X\text{:}\,(r,c)$/$\text{Send}\,X\text{:}\,(r,c)$ name
the peer.}
\label{tab:cholesky_variants}
\end{figure}

\begin{figure}[t]
\centering
\small
\begin{subfigure}[t]{\linewidth}
\centering
\begin{codepanel}
\begin{minipage}{\linewidth}
\raggedright
\noindent
\(\displaystyle
\begin{aligned}
\text{default\_constraint}
&= ((i, j, k),\; i > j > k) \\[0.25em]
(i, i, 0)_{Ls} &\leftarrow A_{ii},
\qquad
(i, j, 0)_{Lg} \leftarrow A_{ij} \\[0.2em]
(i, j, j)_{Lgl}
&\leftarrow \text{TRSM}\big(
  (i, j, j)_{Lc},\;
  (i, j, j)_{Lg}
\big) \\[0.15em]
\color{commhop}
(i, j, k)_{Lgl}
&\leftarrow (i, k, k)_{Lgl}
\qquad (i \ge j) \\[0.15em]
\color{black}
(i, i, k)_{Lgu}
&\leftarrow (i, i, k)_{Lgl} \\[0.15em]
\color{commtree}
(i, j, k)_{Lgu}
&\leftarrow (i - 1, j, k)_{Lgu} \\[0.15em]
\color{black}
(i, j, k)_{Lg}
&\leftarrow \text{GEMM}\big(
  (i, j, k)_{Lgl},\;
  (i, j, k)_{Lgu}, \\
&\qquad\quad
  (i, j, k - 1)_{Lg}
\big) \\[0.15em]
(i, i, k)_{Ls}
&\leftarrow \text{SYRK}\big(
  (i, i, k)_{Lgu},\;
  (i, i, k - 1)_{Ls}
\big) \\[0.15em]
(j, j, j)_{Lc}
&\leftarrow \text{POTRF}\big(
  (j, j, j - 1)_{Ls}
\big) \\[0.15em]
\color{commfact}
(i, j, j)_{Lc}
&\leftarrow (j, j, j)_{Lc}
\end{aligned}
\)
\end{minipage}
\end{codepanel}
\caption{Broadcast $L_{gl}$/$L_c$, systolic $L_{gu}$ (matches Figure~\ref{fig:chol-lowering}).}
\label{fig:chol-dataflow-bcast}
\end{subfigure}

\vspace{0.45em}

\begin{subfigure}[t]{\linewidth}
\centering
\scriptsize
\begin{codepanel}
\begin{minipage}{\linewidth}
\raggedright
\noindent
\(\displaystyle
\begin{aligned}
\color{commhop}
(i, k{+}m, k)_{Lgl}
&= (i, k, k)_{Lgl}
\quad m\in\{1,2,3\}
&& \text{(anchored hops)} \\[0.2em]
\color{commhop}
(i, j, k)_{Lgl}
&= (i, j{-}1, k)_{Lgl}
\quad (j > k{+}3)
&& \text{(relay)} \\[0.2em]
\color{commtree}
(i, j, k)_{Lgu}
&= (i{-}s, j, k)_{Lgu}
\quad (j{+}s \le i < j{+}2s)
&& \text{(tree, } s\in\{1,2,4\}) \\[0.2em]
\color{commfact}
(i, j, j)_{Lc}
&= (i{-}s, j, j)_{Lc}
\quad (j{+}s \le i < j{+}2s)
&& \text{(tree, } s\in\{1,2,4\})
\end{aligned}
\)
\end{minipage}
\end{codepanel}
\caption{Same compute; only communication modified}
\label{fig:chol-dataflow-alt}
\end{subfigure}

\caption{One Cholesky dataflow in Cyclotron, devoid of the statement
ordering, inner-loop nesting, and re-indexing required across the variants
in Figure~\ref{tab:cholesky_variants}. Panel~(a) is the broadcast/systolic
spelling: with $i$ and $j$ spatial and $k$ temporal it lowers to
Figure~\ref{fig:chol-lowering}. Panel~(b) keeps every compute recurrence
and rewrites only $L_{gl}$, $L_{gu}$, and $L_c$---a depth-3 modified ring
along the row and doubling trees down the column.}
\label{fig:tensor_constraints}
\end{figure}

We use distributed, blocked Cholesky decomposition as a motivating example.
Cholesky factors a symmetric positive-definite matrix into a triangular factor and its transpose.
 In a blocked implementation, processors operate on matrix tiles using four dense BLAS/LAPACK kernels:
  POTRF, TRSM, SYRK, and GEMM. Different processors execute different sequences of these operations 
  depending on their position in the processor grid and the current step of the factorization.
  Moreover, these operations depend on values produced elsewhere in the grid, and
  those dependences can be realized using different communication strategies; 
  communication is not simply a uniform pattern of nearest-neighbor shifts. The computation is load--imbalanced, in which critical paths occurring in certain portions of the grid asymmetrically affect work on downstream processors.

Figure~\ref{tab:cholesky_variants} shows two loop-based formulations of blocked Cholesky. 
Even in the serial, non-distributed case, the loop form is coupled to the dependencies among the four kernels. In other words, changing the loop ordering requires more than merely permuting loop headers, as one can for perfectly nested computation such as matrix multiplication. 
It changes loop bounds, the number and placement of inner loops, the ordering of kernel calls, 
and their indexing. In loop--based spatial frameworks, loop dimensions/nesting--order are also used to determine
which parts of the computation execute across processors and which execute over time. 
Changing the space--time mapping can therefore require restructuring the program rather than
simply changing a mapping decision.

Distributing the computation makes the control flow more complex (Figure~\ref{fig:chol-lowering}). The programmer must place sends and receives, add processor- and time-dependent branches,
 construct fan-out send loops, and ensure that every receive has a matching send. 
 Incorrect send/receive pairing can result in indefinite hangs. Changing the communication strategy may require restructuring 
 this control flow even though the underlying computation is unchanged.

Cyclotron instead separates these concerns. The computation is written once in a loopless recurrence form. Figure~\ref{fig:tensor_constraints} shows a Cyclotron recurrence specification of a Cholesky decomposition. It generates the exact code corresponding to Figure~\ref{fig:chol-lowering}, 
while the space--time mapping is specified separately. 
Computation and communication are both expressed as recurrences over iteration-space points, 
so changing the communication strategy  (e.g., from systolic forwarding to broadcast or a tree) changes
only the communication recurrences, not the computation. Figure ~\ref{fig:chol-dataflow-alt} 
shows the same communication recurrences rewritten with a different communication strategy. The communication
of $L_{gu}$ and $L_{c}$ now uses hierarchical doubling tree-broadcasts. 
The communication of $L_{gl}$ is now a modified increasing ring, where it initially broadcasts
to the three nearest neighbors to its right, with the third neighbor and the remaining neighbors
communicating systolically. In the equivalent per-processor imperative program, 
this four-line change more than doubles the code size of the program in Figure~\ref{fig:chol-lowering},
because separate control flow is required for the modified ring (a hybrid systolic-broadcast strategy) and for the doubling tree.

Finally, these recurrence specifications are unordered. Given a space--time mapping and target communication protocol, 
Cyclotron's dependency-management algorithm determines a valid ordering of computation, 
sends, and receives.

\definecolor{codebg}{gray}{0.96}
\section{Notation}
\label{sec:notation}

Recurrences and tensor algebra have traditionally been defined over indexed
tensors \citep{recuma, reptile, vinaithirthan2026compiling, kjolstad2017taco}. Cyclotron augments this with recurrences over \emph{iteration spaces}.
In conventional recurrence and tensor-algebra equations, each tensor
access ($A_{ij}$, or \texttt{A[i][j]}) denotes a logical access to a memory
hierarchy. This model maps well to von Neumann programs on a
single processor, but lacks any mechanism for describing data
that propagates through time or across processors as every data reference is a
tensor access.

Cyclotron's notation describes both computation and communication. First,
\emph{computations} are defined as recurrences over an iteration space
(Section \ref{sec:recurrences}). This way, data may live on the iteration space itself, in addition to living in external tensors. Second, \emph{communication patterns} are specified
in that same language (Section \ref{sec:communication}), as a transfer between two points in the iteration space: a systolic relay, a
broadcast, or a piecewise hybrid of the two is just another recurrence,
distinguished only by the shape of its coordinate expressions. Figure ~\ref{fig:memory-vs-space-intro} illustrates this dichotomy of accessing data from memory versus the iteration space.

\subsection{Defining Recurrences over Iteration Spaces}
\label{sec:recurrences}

\newcommand{\bnfdef}{\mathrel{::=}}
\newcommand{\bnfalt}{\mathrel{\mid}}
\begin{figure}[t]
\centering
\small
\[
\begin{array}{rlcl}
    \multicolumn{4}{l}{\textsf{ConstInt}~n \quad \textsf{Tensor}~t \quad \textsf{IndexVar}~v} \\
    \textsf{Index}        & i  & \bnfdef & v \bnfalt n \bnfalt v+n \\
    \textsf{TensorAccess} & ta & \bnfdef & t_{i_1,\ldots,i_d} \\
    \textsf{IterAccess}   & ia & \bnfdef & (i_1,\ldots,i_d)_t \\
    \textsf{Expr}         & e  & \bnfdef & ta \bnfalt ia \bnfalt \textit{const} \bnfalt \operatorname{func}(e_1,\ldots,e_n) \\
                          &    &         & \bnfalt e+e \bnfalt e\,e \bnfalt \cdots \\
    \textsf{Recurrence}   & r  & \bnfdef & ia \leftarrow e \\
    \textsf{Constraint}   & c  & \bnfdef & i < i \bnfalt i \leq i \bnfalt i = i \bnfalt i \geq i \bnfalt i > i \\
    \textsf{Stmt}         & s  & \bnfdef & r \bnfalt r \;\; c^* \\
    \textsf{Program}      & p  & \bnfdef & s \bnfalt s,p
\end{array}
\]
\caption{Input grammar. \emph{func} does not support recursion.}
\label{fig:grammar}
\end{figure}

In a spatial program, data moves laterally between processors and can be retained
in fast memory across time. If processors are assigned distinct portions of a
program's iteration space, then dataflow across the iteration space corresponds
directly to dataflow between processors, or to dataflow across time.

Cyclotron makes data movement across an iteration space expressible by allowing tensors to be stored on the iteration space, in addition to in memory. Figure~\ref{fig:grammar} shows Cyclotron's grammar. Given a three-tensor $A$
that lives on a three-dimensional iteration space $(i,j,k)$, we write$(i,j,k)_{A}$ to denote the value/tile of $A$ at point $(i,j,k)$ of the iteration space. Figure~\ref{fig:cannonIR} uses this notation to give the
complete dataflow of a systolic matrix multiply in six lines. Movement is encoded by coordinate differences. For example, $(i,j,k)_A \leftarrow (i,\,j{-}1,\,k)_{A}$ propagates the $A$ one step along the second
dimension $j$. This step corresponds to data moving to a neighboring processor if the user declares that dimension to be \emph{spatial}, or to data being retained
on the same processor for the next time step if the user declares the dimension \emph{temporal}. Thus, the same recurrence specification can support different space--time
mappings without changing the equations. In the classical Cannon's algorithm, $i$ and $j$  are spatial and $k$ temporal. This yields systolic movement of $A$ and $B$ and a temporal accumulator for $C$, which is retained in a register from one timestep to another. 

\begin{figure}[t]
\centering
\captionsetup{font=small}
\small

\colorbox{codebg}{%
\begin{minipage}{0.97\columnwidth}
\[
\setlength{\jot}{2pt}
\begin{aligned}
(i,0,k)_{A} &\leftarrow A_{ik}
    && \text{inject $A$} \\
(0,j,k)_{B} &\leftarrow B_{kj}
    && \text{inject $B$} \\
(i,j,0)_{C} &\leftarrow 0
    && \text{init.\ accumulator} \\
(i,j,k)_{A} &\leftarrow (i,j\!-\!1,k)_{A}
    && \text{receive $A$ along $j$} \\
(i,j,k)_{B} &\leftarrow (i\!-\!1,j,k)_{B}
    && \text{receive $B$ along $i$} \\
(i,j,k)_{C} &\leftarrow
    (i,j,k)_{A}(i,j,k)_{B}
    +(i,j,k\!-\!1)_{C}
    && \text{multiply--accum.}
\end{aligned}
\]
\end{minipage}%
}

\caption{Cyclotron specification of a systolic matrix multiplication.}
\label{fig:cannonIR}
\end{figure}

\begin{figure*}[t]
\centering
\includegraphics[width=0.92\textwidth]{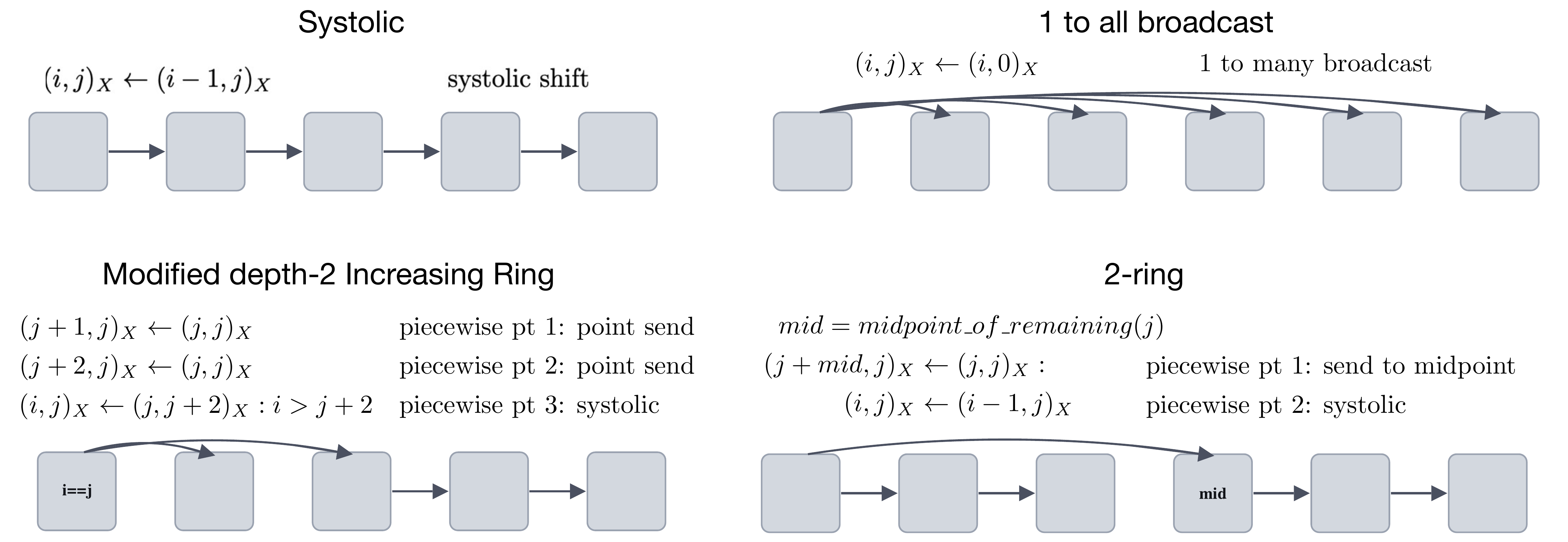}
\caption{Four sample expressible Cyclotron communication patterns, defined over $(i,j)$ with $i$ spatial and $j$ temporal.}
\label{fig:comm-patterns}
\end{figure*}





Computation on the iteration space is expressed with functions over iteration-space values. For example, Figure~\ref{fig:cannonIR} defines matrix multiply-accumulate as
$(i,j,k)_C = (i,j,k)_A (i,j,k)_B + (i,j,k-1)_C$. Because movement is carried in the same coordinate system as computation, it is written in the equations rather than recovered by a separate dependence analysis. In loop-based dataflow languages every data access is a tensor access, so the language can detect that a value must be broadcast across the iteration space but has no notion of how or where the data moves; in Cyclotron the movement and the computation it serves are expressed in one grammar.


Cyclotron specifications are loop-free and unordered. All iteration-space indices are in scope at every recurrence, and no loop order is imposed. The loop-order coupling dissected in Section~\ref{sec:cholesky} (e.g. the $ijk$ vs $kij$ Cholesky loops) is not relevant with recurrences. A separate space--time map linearizes the dependence relation, while remaining separate from the equations to avoid overspecializing programs and to promote program reuse.

\subsection{Iteration Space Boundaries and Subspaces}
Different operations may be defined on subspaces or partitions of the iteration space. Cyclotron supports this through substitutions, which fix indices to constants or other index variables, and constraints, which specify equalities or inequalities over the indexing variables.

Substitutions describe lower-dimensional subspaces of an iteration space. In Figure~\ref{fig:cannonIR}, matrix-multiply operations (i.e., GEMM) are defined over the full rectangular iteration space $(i,j,k)$, while boundary operations are defined on lower-dimensional faces. For example, $(i,0,k)_{A} \leftarrow A_{ik}$ loads tiles from memory where data enters the iteration space, and $(i,j,0)_{C} \leftarrow 0$ seeds the accumulator on the face $k=0$.

Constraints describe subspaces or partitions of a rectangular iteration space. The Cholesky decomposition's inner iteration space is a 3D triangular pyramid defined by $i>j>k$. Figure~\ref{fig:tensor_constraints} shows that a global default constraint can apply to each recurrence, while individual recurrences may specify their own constraints to override it.


Substitutions may also involve other index variables. Cholesky, for example, defines operations on diagonal subspaces such as \(i=j=k\), which form a 1D subset of the 3D iteration space in Figure~\ref{fig:chol-lowering}. We express these by \emph{gimbal-locking} dimensions: substituting one index variable for another, as in \((j,j,j)\). 
Each index variable has a native dimension. Here \(i\) is the first, \(j\) the second, and \(k\) the third. Substituting an index in a non-native dimension equates the indices and locks the dimensions together. Thus \((i,i,i)\), \((j,j,j)\), and \((k,k,k)\) all denote the same diagonal subspace. This differs from loop-based forms, where the specific diagonal index used, such as \(L_{kk}\) versus \(L_{ii}\),  can depend on the surrounding loop structure.


Substitutions, like per-recurrence constraints, override the default constraint when they conflict. If multiple recurrences match a point, the most specific subspace (the one with the most substitutions) takes precedence. For example, \((0,2,2)\) matches a recurrence defined on \((i,j,j)\) over one defined on \((i,j,k)\). Ties are undefined.

\subsection{Specifying Communication}
\label{sec:communication}

Because movement is written as coordinate offsets, a \emph{communication
pattern} is determined by the recurrence's coordinate expressions:
which coordinates carry offsets, and whether the source point is named
relative to the consumer or absolutely. Figure~\ref{fig:comm-patterns}
illustrates four example shapes over an iteration space $(i,j)$ with $i$
spatial and $j$ temporal. We describe four common patterns.

\emph{Systolic relay.} A relative source such as

\begin{equation}
 (i,j)_X \leftarrow (i-1,j)_X 
\end{equation}
forwards values one processor at a time. Offsets need not be unit length; unit offsets denote nearest-neighbor shifts.

\emph{Broadcast.} An absolute or gimbal-locked source such as
\begin{equation}
  (i,\,j)_{X} \leftarrow (j,\,j)_{X} \qquad (i > j),
\end{equation}
causes many consumers indexed by \(i\) to read the same source point where $i==j$. A broadcast will also result from setting the index to a constant, (i.e., $  (i,\,j)_{X} \leftarrow (0,j)_X$), as $(0,j)$ is similarly 1-dimensional. Broadcasts can exist in both space and time; in this example, setting $i$ as a spatial dimension broadcasts $(j,j)_X$ across processors, while naming $i$ in time reuses the value across all later steps.  

\emph{Piecewise hybrids and direct hops} 
Coordinate offsets
may also appear on the \emph{left-hand side}, naming receivers at a fixed
distance from the source. This is used to express hybrids between systolic and broadcast algorithms, in which an initial broadcast (a volley of direct sends) seeds systolic pathways. For example, 
in a \emph{modified increasing
ring} (1rm) shown in Figure ~\ref{fig:comm-patterns}, the source processor sends directly to its first and second right-side neighbors; from the second
neighbor onward, the value is relayed systolically.

We can express it with three pieces:
\begin{align}
  (j{+}1,\,j)_{X} &\leftarrow (j,\,j)_{X} \\
  (j{+}2,\,j)_{X} &\leftarrow (j,\,j)_{X} \\
  (i,\,j)_{X} &\leftarrow (i{-}1,\,j)_{X} \qquad (i > j{+}2)
\end{align}
The result is the first and second-right side neighbors receive
early and can begin computing immediately, while the source pays for
only two sends rather than a broadcast. This is especially useful in solvers and decompositions, where early processors in a row gate more work than later processors, and benefit from starting early.

Each point of the iteration space matches one of these pieces. The direct hops are more specific than the guarded relay and therefore take precedence; the compiler lowers the pieces to corresponding control-flow guards (i.e., if statements). The guards are ordered such that at runtime, each processor at each timestep binds to the most specific piece. In this case, because the first two pieces are more specific (written in terms of 1 variable, thus exhibiting gimbal lock), its guards are evaluated first. Adding \((j+3,j)_X \leftarrow (j,j)_X\) and shifting the relay guard to \(i>j+3\) deepens the direct-send prefix to depth three, and similarly for larger depths, trading source fan-out against pipeline fill time.




\emph{Two-rings.} A direct hop need not target a fixed offset. In
an \emph{increasing-2-ring}, the source sends to its nearest neighbor and to the midpoint of
the remaining processors, creating two concurrent systolic relay chains. With $m = \lceil (P - j)/2 \rceil$ the midpoint of the
processors yet to be reached,
\begin{align}
  (j{+}m,\,j)_{X} &\leftarrow (j,\,j)_{X} \\
  (i,\,j)_{X} &\leftarrow (i{-}1,\,j)_{X} \qquad (i \neq j{+}m)
\end{align}
As shown in Figure~\ref{fig:comm-patterns}, the two sends seed separate systolic rings, roughly halving the relay depth at the cost of one additional source send. Because \(m\) depends on the iteration-space coordinate, the same form naturally generalizes to other multi-ring strategies.


These patterns differ only in how guarded recurrences name their sources and receivers. By composing them, Cyclotron can express strategies such as split rings (where systolic communication starts at a source processor and proceeds in both directions), doubling broadcast trees, and heterogeneous piecewise combinations, rather than restricting users to a fixed set of communication presets.


\section{Compilation and Dependency Analysis}
\label{sec:compilation}
Compilation lowers a recurrence system and space--time mapping into one per-processor program. Algorithm~\ref{alg:compile} shows the full compilation pipeline. Spatial dimensions are bound to processor coordinates and temporal dimensions become loops. The dependency-management pass uses the space--time mapping to classify recurrence dependences as live or stale. It then topologically orders the live dependences, and then emits matching communication. Subsequent passes make communication synchronization-safe and optionally apply block-cyclic mapping and static deduplication.

\begin{algorithm}[t]
\caption{Compiling recurrence to a distributed program.}
\label{alg:compile}
\begin{algorithmic}[1]
\Require recurrence system $\mathcal{R}$; space-time annotation $\sigma$
\Ensure one guarded program run on every processor
\State emit a \textbf{loop} over the temporal dimensions of $\mathcal{R}$
\State bind spatial coordinates from the processor rank
\State $G \gets \textsc{DependenceGraph}(\mathcal{R})$
\ForAll{edges $e \in G$}
  \If{$\textsc{Offset}(e)$ lies purely along a temporal dim}
    \State mark $e$ \emph{stale} \Comment{ resident in a register}
  \Else
    \State mark $e$ \emph{live} \Comment{produced this timestep}
  \EndIf
\EndFor
\State $\textit{order} \gets \textsc{TopoSort}(\text{live edges of } G)$
  \Comment{emission order}
\ForAll{tensor $T$ in \textit{order}}
  \State emit \textbf{if} guards for $T$: from substitutions/constraints
  \State emit $T$'s guarded block: receives, reads, kernel calls
  \State emit \textbf{endIf} 
\EndFor
\State $\textsc{DeducePushes}(\mathcal{R})$ emits \textbf{Send}s
\State \textbf{emitEndLoop} \Comment{close the temporal loop body}
\If{block-cyclic mapping requested} \Comment{optional}
  \State $\textsc{BlockCyclicFold}$: wrap program in virtual-processor loops; add modulo indexing and guards
\EndIf
\State $\textsc{SpecializeProgram}$: specialize guards/loops per rank; drop redundant sends/recvs \Comment{dedup communication}
\State \Return interpreted per-processor program
\end{algorithmic}
\label{fig:algo}
\end{algorithm}


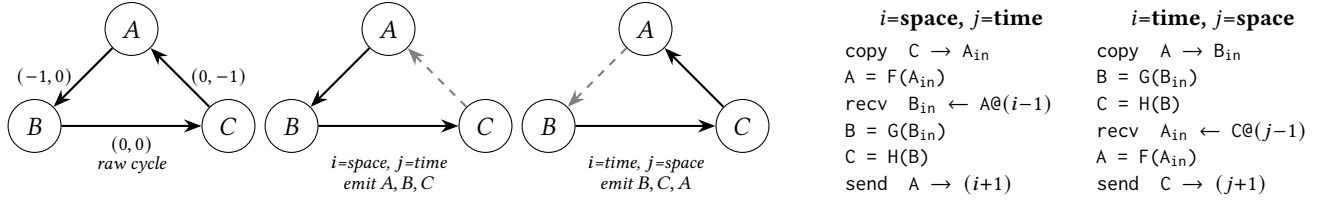
\begin{figure*}[t]
\centering
\begin{minipage}[c]{0.60\textwidth}
\centering
\begin{tikzpicture}[
    reg/.style={circle, draw, minimum size=7mm, inner sep=0pt},
    dep/.style={-{Stealth}, thick},
    stale/.style={-{Stealth}, thick, dashed, gray},
    node distance=11mm]
  \node[reg] (A1) {$A$};
  \node[reg] (B1) [below left=of A1] {$B$};
  \node[reg] (C1) [below right=of A1] {$C$};
  \draw[dep] (A1) -- node[left, font=\scriptsize] {$(-1,0)$} (B1);
  \draw[dep] (B1) -- node[below, font=\scriptsize] {$(0,0)$} (C1);
  \draw[dep] (C1) -- node[right, font=\scriptsize] {$(0,-1)$} (A1);
  \node[below=12mm of A1, font=\scriptsize\itshape] {raw cycle};

  \begin{scope}[xshift=34mm]
  \node[reg] (A2) {$A$};
  \node[reg] (B2) [below left=of A2] {$B$};
  \node[reg] (C2) [below right=of A2] {$C$};
  \draw[dep]   (A2) -- (B2);
  \draw[dep]   (B2) -- (C2);
  \draw[stale] (C2) -- (A2);
  \node[below=12mm of A2, font=\scriptsize\itshape, align=center]
    {$i$=space, $j$=time\\ emit $A,B,C$};
  \end{scope}

  \begin{scope}[xshift=68mm]
  \node[reg] (A3) {$A$};
  \node[reg] (B3) [below left=of A3] {$B$};
  \node[reg] (C3) [below right=of A3] {$C$};
  \draw[stale] (A3) -- (B3);
  \draw[dep]   (B3) -- (C3);
  \draw[dep]   (C3) -- (A3);
  \node[below=12mm of A3, font=\scriptsize\itshape, align=center]
    {$i$=time, $j$=space\\ emit $B,C,A$};
  \end{scope}
\end{tikzpicture}
\end{minipage}\hfill
\begin{minipage}[c]{0.36\textwidth}
\centering
\small
\begin{minipage}[t]{0.48\linewidth}
\centering
\textbf{$i$=space, $j$=time}
\begin{flushleft}\ttfamily\footnotesize
copy\ \ C $\to$ A\textsubscript{in}\\
A = F(A\textsubscript{in})\\
recv\ \ B\textsubscript{in} $\leftarrow$ A@$(i{-}1)$\\
B = G(B\textsubscript{in})\\
C = H(B)\\
send\ \ A $\to$ $(i{+}1)$
\end{flushleft}
\end{minipage}%
\hfill
\begin{minipage}[t]{0.48\linewidth}
\centering
\textbf{$i$=time, $j$=space}
\begin{flushleft}\ttfamily\footnotesize
copy\ \ A $\to$ B\textsubscript{in}\\
B = G(B\textsubscript{in})\\
C = H(B)\\
recv\ \ A\textsubscript{in} $\leftarrow$ C@$(j{-}1)$\\
A = F(A\textsubscript{in})\\
send\ \ C $\to$ $(j{+}1)$
\end{flushleft}
\end{minipage}
\end{minipage}
\caption{Left: the dependence graph of
Equations~\ref{eq:abc-a}--\ref{eq:abc-c} is a cycle (edges point from
producer to consumer and are labeled with their offsets). The space--time
mapping determines which edge is \emph{stale} (dashed): a purely temporal
offset refers to a value already resident from a previous timestep. Cutting
the stale edge leaves an acyclic graph. The topological order determines emitted statement order. Right: the ordering of statements changes under the two mappings. }
\label{fig:dep-cycle}
\label{fig:emitted-ir}
\end{figure*}

\subsection{The shape of generated code}

The generated program is the same on every processor: a single loop over the
temporal dimension, whose body is a sequence of guarded statements. Each guard
is an \texttt{if} over the processor's spatial coordinates and the current
value of the temporal loop variable. The guards come from two sources: the
case structure implied by gimbal-locking (a statement written on a boundary
face such as $(i,i,k)$ only applies where $i=j$), and constraints the user
supplies directly (such as $i > j > k$). They are necessary because both the
temporal loop and the spatial processor grid range over the entire
iteration space: every processor reaches every timestep, but only some
$(\text{processor},\, \text{time})$ points lie inside a given statement's
region. The guard determines whether the current processor and timestep belong to the statement's guarded region.

Within one iteration of the temporal loop, each tensor of the recurrence
system owns one guarded block. A block contains everything that produces that
tensor's value for the current timestep: a receive if the value arrives from a
neighboring processor, a register read if it is already resident from a prior
timestep, and the kernel call that computes it. Sends are placed at the end of
the loop body, after all values for the step have been produced. The compiler must therefore determine an emission order for these blocks that respects their data dependencies.

\subsection{Breaking cyclic recurrence dependencies}
\label{sec:live-stale}

The order matters, and it cannot be read off the source program order, because the
dependence graph may be cyclic. Consider the following three-tensor system, which we use as a
running example (Figure~\ref{fig:dep-cycle}):
\begin{align}
  (i,j)_{A} &= F\big((i,\,j{-}1)_{C}\big) \label{eq:abc-a}\\
  (i,j)_{B} &= G\big((i{-}1,\,j)_{A}\big) \label{eq:abc-b}\\
  (i,j)_{C} &= H\big((i,\,j)_{B}\big)     \label{eq:abc-c}
\end{align}
$B$ depends on $A$, $C$ on $B$, and $A$ on $C$: a cycle
$A \to B \to C \to A$. $G$, $H$, and $F$ are functions. Taken at face value, the cyclic dependencies admit no valid emission order for the current timestep; an arbitrary order can produce incorrect code.

The cycle is broken by a space--time mapping. Once one dimension is declared
temporal, any dependence whose offset lies purely along the temporal
dimension refers to a value from a previous timestep. That value
already exists,  computed in an earlier iteration of the temporal loop
and is sitting in a register. It does not need to be produced during the
current step, so it imposes no ordering constraint on the current step's
code. We call such dependence edges \emph{stale}, and we remove them from the
ordering graph. The read itself is not removed; it is lowered to a
register read. Remaining edges, whose values are  produced during the current step,
are \emph{live}. Deleting stale edges leaves an acyclic graph. A
topological sort of the live edges yields a valid emission order.

In the example, map $i$ to space and $j$ to time. The dependence
$(i,j)_A \leftarrow (i,\,j{-}1)_{C}$ has offset $(0,-1)$: purely temporal, thus 
stale; the last step's $C$ is already in a register. The dependence
$(i,j)_{B} \leftarrow (i{-}1,\,j)_{A}$ has offset $(-1,0)$ and is purely spatial. It is hence live and this step's $A$, arriving from the neighboring processor. $(i,j)_C \leftarrow (i,j)_{B}$ has no offset and is live under any
mapping. Cutting the stale edge leaves the chain $A \to B \to C$, and the
emitted order is $A,\,B,\,C$.

The flipped mapping, with $i$ to time and $j$ to space, results in a different emitted order. Every offset keeps its
spelling but changes its meaning. The dependence $(i,j)_A \leftarrow (i,\,j{-}1)_{C}$  is now
\emph{spatial}, and the dependence  $(i,j)_B \leftarrow (i{-}1,\,j)_{A}$ is now \emph{temporal} and the source of the cut. 
The surviving chain is $B \to C \to A$, different than the original A,B,C order.

This reordering is necessary for correctness, not a stylistic choice. Suppose
we kept the source order $A,\,B,\,C$ under the flipped mapping. $B$'s
dependence on $(i{-}1,\,j)_{A}$ is stale: it must read the value of register
$A$ left over from the previous timestep. But the block for $A$ has
already run and overwritten that register with the current step's
value. $B$ silently consumes a value one timestep too new, resulting in incorrect output. The stale value must be read \emph{before} the operation that clobbers its register is emitted.

The topological sort in lines 3-11 of Algorithm~\ref{fig:algo} guarantees this. With $i$ mapped in time,  the stale edge $A \to B$ is deleted, its consumer $B$ loses its only incoming edge and
becomes a leaf of the ordering graph, so it is emitted
first, before the block that overwrites $A$. This is not a coincidence of
the example. The stale edge closes the cycle, so the surviving live edges
form a path from the stale edge's consumer ($B$) back around to its producer
($A$); any topological order of that path places the consumer before the
producer. In other words, sorting on live edges alone also enforces the
read-before-clobber constraint of the stale edge, for free.

The same structure appears at full scale in Cholesky. In this case the cycle
is $L_s \to L_c \to L_{gl} \to L_{gu} \to L_s$. The cut again changes with
the space--time mapping. The resulting topological orders correspond to the different compute orders required by the loop variants in Figure~\ref{tab:cholesky_variants}. 
The topological algorithm produces those orderings directly. Fixed traversals of the
source program (depth-first from any starting tensor) cannot produce correct orders for all mappings, because it must cut the cycle, and the correct cut is determined by the space--time mapping, not program order.


\subsection{Emitting Communication}
\label{sec:emitting-comm}

Once blocks are  lowered, cross-processor dependences are emitted as  receives. If $i$ is spatial, $(i,j)_X\leftarrow(i-1,j)_X$ means processor $i$ receives $X$ from processor $i-1$. The user does not write the corresponding send; the compiler inverts the relation and inserts a send from processor $i-1$ to $i$.

\begin{table*}[t]
\centering
\caption{Send synthesis in \textsc{DeducePushes}. Each receive
(\emph{dst}\,$\leftarrow$\,\emph{src}) is classified by its coordinate
and matched with a send.} 
\label{tab:send-taxonomy}
\small
\renewcommand{\arraystretch}{1.35}
\begin{tabular}{@{}lllll@{}}
\toprule
\textbf{Pattern} & \textbf{Receive \& window (dst\,$\leftarrow$\,src)} &
\textbf{Relay test} & \textbf{Originates at} &
\textbf{Send \& guard (src\,$\rightarrow$\,dst)} \\
\midrule
Temporal carry & $(i,j)\leftarrow(i,j{-}1)$ & --- & --- (register persists) & --- none; value stays resident \\
Systolic relay & $(i,j)\leftarrow(i{-}1,j),\ \ \mathit{lo}\le i<\mathit{hi}$ & $\mathit{hi}{-}\mathit{lo}>\text{stride}$ & receiver (each PE forwards) & $(i,j)\rightarrow(i{+}1,j),\ \ \mathit{lo}\le i{+}1<\mathit{hi}$ \\
Scatter (non-relay) & $(i,j)\leftarrow(i{-}\text{stride},j),\ \ \mathit{lo}\le i<\mathit{hi}$ & $\mathit{hi}{-}\mathit{lo}\le\text{stride}$ & each holder, one stride up & $(i,j)\rightarrow(i{+}\text{stride},j),\ \ \mathit{lo}\le i{+}\text{stride}<\mathit{hi}$ \\
Anchored       & $(j{+}1,j)\leftarrow(j,j)$ & --- & source PE & $(j,j)\rightarrow(j{+}1,j)$ \\
Broadcast      & $(i,j,k)\leftarrow(i,k,k)$ & --- & source PE (fan-out) & $(i,k,k)\rightarrow(i,j,k)\ \ \forall j$ \\
\bottomrule
\end{tabular}
\end{table*}

Table ~\ref{tab:send-taxonomy} shows how sends are synthesized, depending on the dependence.
Temporal dependencies require no send because the value remains locally resident.  Anchored receives such as $(j+1,j)_X \leftarrow (j,j)_X$ name the destination directly in terms of the source coordinate, so the source emits a send to $j+1$. Broadcasts such as $(i,j)_X \leftarrow (j,j)_X$  map many receivers $i$ to the same source $j$, so source $j$ emits sends to each receiver. Same-base offset receives like $(i,j)\leftarrow (i-1,j)$ require one additional test to check if receivers also forward the value (producing a systolic relay), or only receive.  

Consider a receive $(i,j)_X\leftarrow(i-stride,j)_X$  defined over the window $low \leq i < hi$. If $stride< hi-low$, then some receivers read from another point within the window, meaning some receivers also forward the value. If $stride \geq hi-low$, every sender remains outside the window, so receivers do not forward; this produces a scatter. Cyclotron uses scatters to construct recursive-doubling tree broadcast. 

The user never writes sends or send-side constraints. Cyclotron simply copies the receiver's if guards onto the synthesized send, ensuring the sender fires only when its destination is a receiver. Duplicate sends are suppressed when multiple rules would serve the same destination.

\subsection{Minimizing communication}
\label{sec:block-cyclic}


Triangular iteration spaces like Cholesky’s are load imbalanced under a one-tile-per-processor mapping. Cyclotron therefore supports a 2D block-cyclic decomposition, which maps the virtual processor grid onto a smaller physical grid in round-robin fashion. This both balances work and exposes redundant communication. If several virtual processors on the same physical processor require the same tile from another processor, only one network receive is necessary and the remaining consumers/virtual processors  can reuse the same local copy. Cyclotron applies this decomposition in two phases. The first is a post-processing pass that wraps each processor program with loops over its resident virtual processors and inserts the required modulo and guard logic. Afterwards, the generated code is specialized to each physical processor; control flow is removed, as well as redundant sends, receives, and guards. The resulting per-processor programs remain small because each generated operation corresponds to a coarse-grained tiled computation. An alternative is to eliminate redundant communication with a runtime cache, but this introduces additional runtime overhead.

\section{System and Implementation}

The Cyclotron front-end language and compiler are implemented in Python, while the runtimes that execute generated programs are native C++ and Rust. Cyclotron has three backends: a C++ backend for CPU clusters, a C++ backend for GPU clusters, and a Rust backend built with DAM \citep{zhang2024dataflow}, a framework for modeling dataflow architectures in which PEs communicate over channels. Each backend interprets the same portable Protobuf program, allowing a generated executable to run unchanged across backends. The Rust backend is used primarily for simulation. It reproduces the communication behavior of the cluster backends, including simulated MPI protocol behavior, without requiring payloads. This allows Cyclotron to detect indefinite hangs and model communication and kernel latencies without executing on a cluster. The simulator can also execute with real payloads and produce correct results, although this mode is disabled by default because it runs on a single processor. The Rust backend can be used to simulate spatial-processors composed of interlinked PEs; this exploration is future work.


\section{Evaluation}
\label{sec:eval}
We evaluate performance on CPU and GPU clusters. Both are hosted on Perlmutter, a supercomputer at Lawrence Berkeley National Laboratory. GPU experiments use 16 A100 GPUs, four per node, connected by NVLink within a node and HPE Slingshot 11 across nodes. CPU experiments use 16 Perlmutter nodes connected by Slingshot. 
We compare against Cray ScaLAPACK, SLATE, and cuSolverMp using matched precision, tile sizes, BLAS/LAPACK kernels, thread counts, and MPI communication. Each configuration uses two warmups and four measured trials; we report the best runtime.

\subsection{Comparing Communication Strategies}
\label{sec:eval-comm}

\begin{figure*}[t]
\centering
\begin{subfigure}{0.49\textwidth}
\centering
\includegraphics[width=\linewidth]{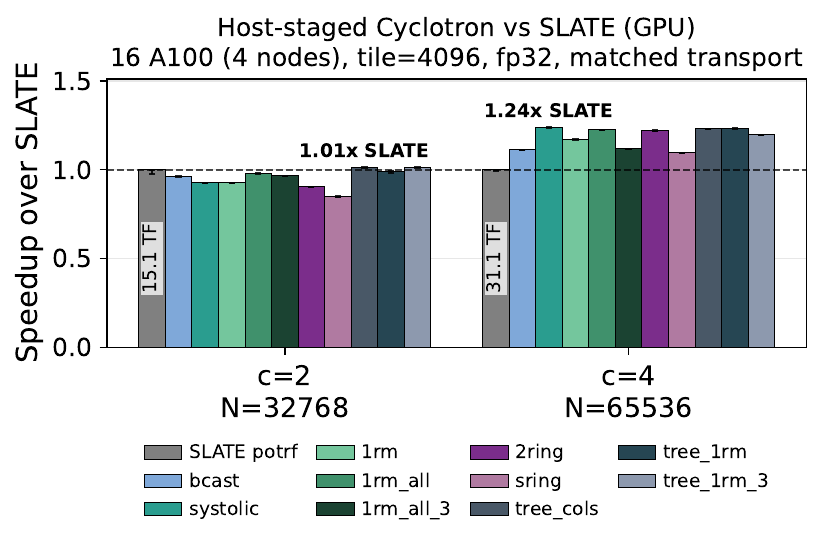}
\label{fig:gpubc-slate-t2048}
\end{subfigure}
\hfill
\begin{subfigure}{0.49\textwidth}
\centering
\includegraphics[width=\linewidth]{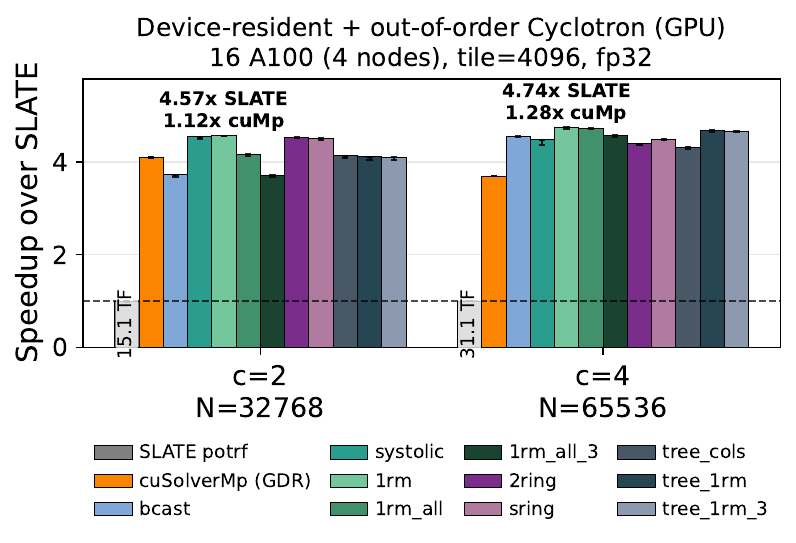}
\label{fig:gpubc-slate-t4096}
\end{subfigure}
\caption{Block-cyclic Cholesky on four GPU nodes ($16$ A100s, 1 rank/GPU,
so $P{=}4$): 
Left: speedup over SLATE, with host-staged communication.  Right: speedup over Nvidia cuSolverMp and SLATE, with device-device direct communication.
tilesize=$4096$, with block-cyclic factor $c{=}2$ and $c{=}4$ (right). Each
process owns $c{\times}c$ tiles, so $N=P\,c\cdot\text{tile}$; SLATE and
cuSolverMp use the same block size ($=$tile) and the same $c$. Bars are the
fastest steady cycle; whiskers reach  slowest.}
\label{fig:gpubc-slate}
\end{figure*}

\begin{figure}[t]
\centering
\includegraphics[width=\columnwidth]{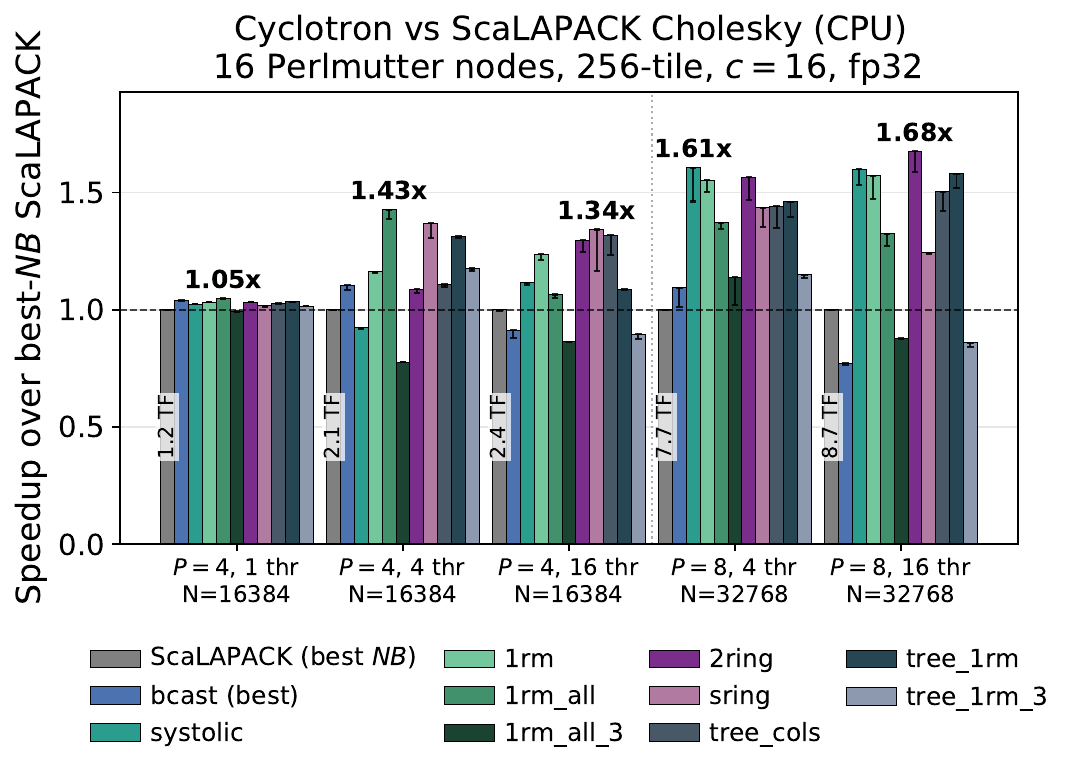}
\caption{Cyclotron Cholesky speedup over ScaLAPACK on 16 CPU nodes. \(P\times P\) ranks, tilesize=256, 16 tiles/rank; “thr” denotes threads/rank. Whiskers show run-time variation}
\label{fig:bc16av}
\end{figure}

\begin{figure}[t]
\centering
\includegraphics[width=\columnwidth]{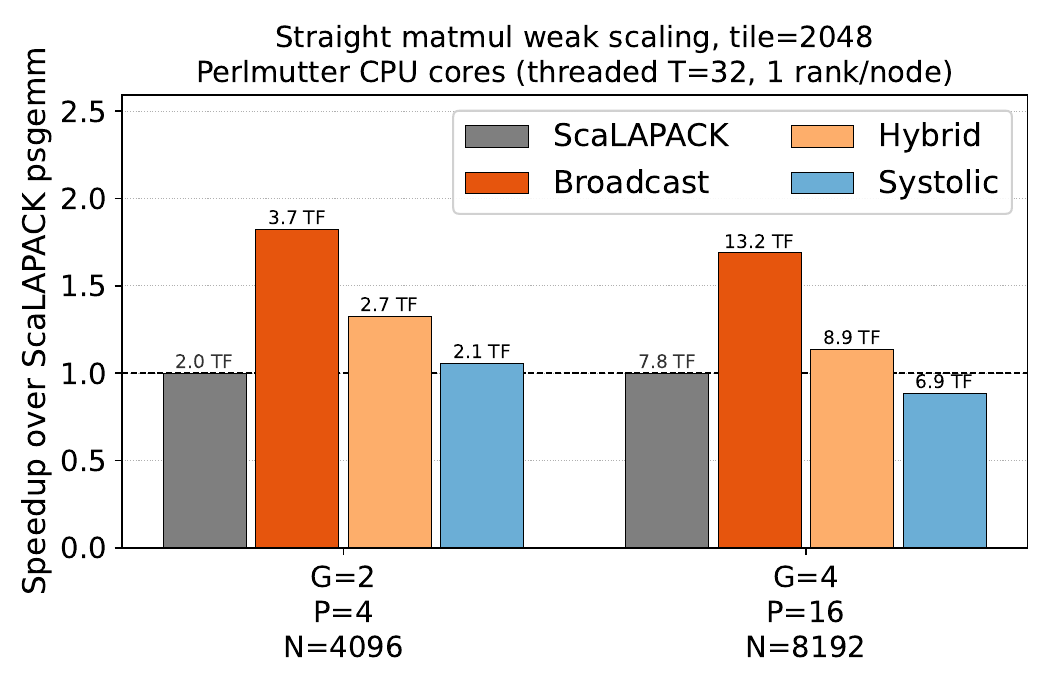}
\caption{Cyclotron GEMM vs. ScaLAPACK on 4 and 16 CPU nodes, with systolic, broadcast, and hybrid schedules}
\label{fig:matmulRes}
\end{figure}

For a specific program, the best communication schedule is rarely fixed; the fastest one will depend on factors including the network topology and processor architecture (CPUs vs GPUs), in addition to tile size, tiles per rank, and input size. The relative performance of schedules also changes; we observed a single schedule can perform over twice as fast as an alternative on CPUs, while the alternative often performs faster than the other on GPUs.

We compare a sample set of eleven Cyclotron Cholesky variants that differ only in communication strategy, including broadcasts, systolic relays, 2-rings, bidirectional split-rings (sring), modified rings (1rm), and trees, against ScaLAPACK and SLATE. We use an $ij$ spatial mapping here because it enables communication deduplication, which is fundamentally not possible in other Cholesky spatial mappings.


\paragraph{CPU Results} Figure~\ref{fig:bc16av} compares 11 Cyclotron Cholesky communication schedules against ScaLAPACK on 16 CPU nodes using \(P\times P\) grids, \(P\in\{4,8\}\), with 4 or 16 threads/rank. We also include an effectively serial control (\(P=4\), 1 thread/rank), where communication choice has little effect and all Cyclotron schedules perform similarly to ScaLAPACK.

ScaLAPACK's block size \(NB\) was swept; the best was \(256\) in all cases, which we use as the baseline. Cyclotron also uses \(NB=256\), giving \(c=16\). As per-rank computation becomes more parallel, communication becomes increasingly important: some schedules perform worse than ScaLAPACK, while the best Cyclotron schedule achieves \(1.34\)--\(1.68\times\) speedup across the remaining configurations. The winning communication strategy also changes with the configuration, and schedules that perform well in one regime can perform poorly in another. This demonstrates that communication policy becomes a first-order performance choice once computation is sufficiently optimized.
The choice of communication policy is also heavily machine dependent. On Perlmutter's 16 nodes, the broadcast policy performed poorly. However, on a separate 4-node  machine (with nodes connected via InfiniBand), the broadcast schedule was universally best. We postulate this is due to the 4-nodes being densely connected, making 1 to all broadcasts practical.

\paragraph{GPU Results.}
Figure~\ref{fig:gpubc-slate} (left) compares Cyclotron against SLATE using host-staged communication on 16 A100 GPUs. We use tile size 4096 and evaluate block-cyclic factors \(c=2\) and \(c=4\). We use host staging here because SLATE (on Perlmutter) does not support GPU-direct communication, giving both systems the same communication path. Under this matched transport, Cyclotron is competitive with SLATE at \(c=2\) and reaches up to \(1.24\times\) SLATE at \(c=4\). The relatively smaller spread among Cyclotron's communication schedules contrasts with the CPU results; the importance of communication strategy is architecture dependent.

Figure~\ref{fig:gpubc-slate} (right) evaluates Cyclotron on 16 A100s with device-resident GPU-direct communication and out-of-order execution, and compares against Nvidia cuSolverMp, which likewise uses GPU-direct communication. Cyclotron reaches \(4.57\times\) and \(4.74\times\) SLATE, and \(1.12\times\) and \(1.28\times\) cuSolverMp. 

The difference with cuSolverMp appears to come largely from execution scheduling. cuSolverMp is closed source, but profiling shows that its communication is performed using NCCL broadcast collectives, with communication and computation proceeding in bulk-synchronous stages. Cyclotron instead emits dependency-driven code that naturally overlaps communication and computation, and out-of-order execution further increases this overlap. 


\paragraph{CPU Matrix Multiply.} We also evaluate three different matrix multiply schedules against ScaLAPACK on our 16-node CPU backend. We compare systolic, broadcast, and hybrid schedules (broadcast in 1 dimension, systolic in the other). Figure ~\ref{fig:matmulRes} shows the broadcast variant working best.

\subsection{Generating Different Spatial Mappings}
\label{sec:eval-schedules}

\begin{figure}[t]
\centering
\includegraphics[width=\columnwidth]{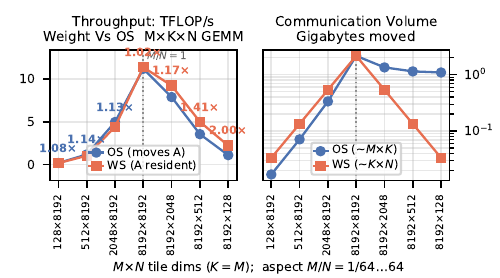}
\caption{Weight-stationary (\(ik\)-spatial) versus output-stationary (\(ij\)-spatial) GEMM on 16 CPU nodes. As \(N\) decreases relative to \(M=K\), weight-stationary increasingly wins by keeping the larger \(A\) tiles stationary.}
\label{fig:WSvsOS}
\end{figure}

The space--time annotation of Section~\ref{sec:notation} is a retag, not a
rewrite. The same blocked triangular-solve recurrences produce two complementary
wavefronts depending on which index is spatial: an $i$-spatial  pipelines the solution $X$, while a $k$-spatial pipelines the right-hand side $B$. Each leaves a  region of processors idle while waiting for its wavefront, but these idle regions are complementary. One can therefore interleave the two mappings by changing only the space--time annotation between successive right-hand-side chunks.



\begin{table}[t]
\centering
\caption{Blocked \textsc{trsm} throughput on four CPU nodes
(Perlmutter), tile $1024\times1024$, $1024$ right-hand sides).
Entries are GFLOPS/proc. Overlap is ping-pong time against the sum of the
two isolated systolic times. ScaLAPACK is \texttt{pstrsm} on the same
$P{\times}1$ grid and the same number of solves (20)}
\label{tab:trsm-pingpong}
\small
\begin{tabular}{@{}l r r@{}}
\toprule
 & $P=4$ (1/node) & $P=8$ (2/node) \\
\midrule
$i$-spatial systolic & 54.37 & 49.95 \\
$k$-spatial systolic & 54.54 & 50.36 \\
broadcast & 55.70 & 51.86 \\
modified ring (1rM) & 55.39 & 50.75 \\
increasing 2-ring & 55.92 & 49.89 \\
ping-pong systolic & \textbf{93.65} & \textbf{91.98} \\
ScaLAPACK \texttt{pstrsm} & 47.55 & 44.33 \\
\midrule
overlap speedup & $1.72\times$  & $1.83\times$ \\
vs.\ ScaLAPACK & $1.97\times$ & $2.07\times$ \\
\bottomrule
\end{tabular}
\end{table}

Table~\ref{tab:trsm-pingpong} evaluates this ``ping-pong'' schedule. Even-numbered chunks use one spatial mapping and odd-numbered chunks use the other, allowing work from one wavefront to fill much of the other's pipeline imbalance. In isolation, the Cyclotron variants perform similarly: changing the spatial dimension or replacing a systolic relay with a broadcast, modified ring, or 2-ring has little effect on the critical path of a single solve. Interleaving the two complementary systolic mappings, however, improves throughput by \(1.72\times\) at \(P=4\) and \(1.83\times\) at \(P=8\) relative to running them back-to-back, reaching \(1.97\times\) and \(2.07\times\) the throughput of ScaLAPACK. The ping-pong throughput remains similar across the two placements (93.65 vs.\ 91.98 GFLOPS/proc), indicating that the benefit does not arise merely from packing multiple ranks onto a node.

Figure~\ref{fig:WSvsOS} compares two spatial mappings of the same systolic matrix multiply and shows that the better mapping depends on tile shape. We compute $C=AB$ on a $4\times 4$ processor grid, where $A$ is $M\times K$, $B$ is $K\times N$, and $C$ is $M\times N$. We fix $M=K$ (square $A$ tiles) and sweep $N$. When $M=K=N$, both dataflows perform equivalently. As $N$ shrinks, $A$ becomes much larger than $B$ and $C$: the output-stationary mapping moves $A$ and $B$, whereas the weight-stationary mapping keeps the larger $A$ tiles fixed and moves only the smaller $B$ and $C$ tiles. Consequently, the weight-stationary mapping outperforms output-stationary as $N$ decreases.


\subsection{Fusion}
\label{sec:eval-fusion}

\begin{table}[t]
\centering
\caption{Fused systolic Cholesky\,+\,TRSM versus ScaLAPACK
\texttt{pspotrf} then \texttt{pstrsm} ($N=4096$, $4\times4$ grid,
$16$ ranks). Times in milliseconds. Hidden is the fraction of our
standalone 2-D solve absorbed by fusion,
$1-(t_{\mathrm{fused}}-t_{\mathrm{chol}})/t_{\mathrm{trsm}}$.}
\label{tab:chol-fusion}
\small
\begin{tabular}{@{}l rrr rrr@{}}
\toprule
 & \multicolumn{3}{c}{$c{=}1$ (tile $1024$)} & \multicolumn{3}{c}{$c{=}4$ (block-cyclic, tile $256$)} \\
\cmidrule(lr){2-4} \cmidrule(lr){5-7}
$n_{\mathrm{rhs}}$ & fused & hidden & vs.\ scala
                   & fused & hidden & vs.\ scala \\
\midrule
$256$ & $180$ ms & $90\%$ & $1.17\times$ & $48$ ms & $23\%$ & $1.62\times$ \\
$512$ & $182$ ms & $90\%$ & $1.34\times$ & $58$ ms & $18\%$ & $1.30\times$ \\
\bottomrule
\end{tabular}
\end{table}

Each PE in a Cyclotron program eagerly executes an operation as soon as its dependencies are satisfied. Consequently, running subsequent kernels does not introduce a global barrier between them: operations from the second kernel can begin before the first kernel has completed across all PEs. The preceding ping-pong TRSM experiment demonstrated this effect across successive instances of the same kernel, where pipeline imbalances in adjacent solves overlap naturally without an explicit user-directed fusion primitive.

The same mechanism applies across different kernels; we also execute a Cholesky followed by a multi-right-hand-side triangular solve. Table~\ref{tab:chol-fusion} reports this experiment on four CPU nodes in both block-cyclic and non-block-cyclic regimes. In the non-block-cyclic configuration, Cholesky is load-imbalanced, and processors in earlier rows and columns of the processor grid finish their assigned factorization work substantially earlier than others. Because Cyclotron executes operations as soon as their dependencies are satisfied, portions of the subsequent triangular solve can begin on these processors while Cholesky is still running elsewhere. This combination of load imbalance and dependency-driven fusion hides about 90\% of the standalone solve time, allowing the solve to execute nearly for free. In the block-cyclic configuration, work is more evenly balanced across processors, leaving less idle time to exploit; fusion therefore hides only 18--23\% of the solve, while still outperforming sequential ScaLAPACK Cholesky followed by TRSM by \(1.30\)--\(1.62\times\).

\section{Related Work}
\label{sec:related}

Systolic arrays were introduced by Kung and Leiserson~\citep{kung1979systolic,kung1982why}. Rajopadhye, Fujimoto, and Purushothaman~\citep{rajopadhye1990synthesizing, sanjay-fst-tcs, sanjay-thesis}, as well as Quinton \citep{Quinton84c, quinton1989mapping} established how to synthesize them from uniform and affine recurrence equations. 
There is a long history of dataflow compilers based on the polyhedral and affine models. AutoSA~\citep{autosa} and PolySA~\citep{cong2018polysa} use the polyhedral model to find unit-length read distances among loop nests and then compile the equations to dataflow. They primarily target matrix multiply and convolution, and do not handle recurrences (equations with dependencies) beyond the LU decomposition, whose dataflow resembles matrix multiply. 
Lam did early work on systolic-array compilation, while Lam and Wolfe studied locality optimization \citep{lam1987systolic, lam2004data}.


AlphaZ~\citep{alphaz, sanjay-yuki-lcpc2012, sanjay-yuki-TR13-105} is the closest equational system, instantiating recurrence equations on distributed-memory machines. Users explicitly provide space--time maps, including statement-ordering dimensions that Cyclotron instead infers. Its distributed backend emits nearest-neighbor systolic halo exchanges using blocking \texttt{MPI_Send} and \texttt{MPI_Recv}, whereas Cyclotron expresses broadcasts, relays, and piecewise rings directly in the recurrence system. Legion~\citep{legion} also derives communication from programmer-declared data dependencies, but does so dynamically at runtime rather than statically. Its programs also explicitly order per-timestep kernel launches, whereas Cyclotron derives this ordering from the space--time mapping. Thus, Cyclotron's ordering algorithm is complementary to Legion and could sit above it.

Our IR is an extension and re-imagining of the \emph{assignment-free} notation from the \emph{Algorithms of Informatics} textbook by Gyires~\citep{gyires_systolic}. Their notation is array-style, similar to Halide's input language~\citep{ragan2013halide} (e.g.\ $A(i,j,k)$, as opposed to our $(ijk)_{A}$). Halide's notation is adapted by the Stellar~\citep{genc2024stellar} and Gemmini~\citep{genc2021gemmini} work for designing matrix-multiply systolic arrays. Cyclotron extends this style of notation to express multiple communication patterns, combinations of them, and gimbal-locked indexing. Calyx~\citep{nigam2021calyx} and Dahlia~\citep{nigam2020dahlia} provide IRs for synthesizing accelerators with explicit control and memory pipelines, but do not reason about affine recurrences or induce space--time schedules. The Wavefront Array Processor~\citep{wavefrontLanguage} was a programmable systolic array whose language resembled the per-PE programs Cyclotron generates.

DISTAL~\citep{distal} is a tensor-algebra compiler, built on TACO~\citep{kjolstad2017taco}, that generates distributed matrix-multiplication kernels. DISTAL separates tensor distribution from computation scheduling, with communication inferred from these choices; it supports only output-stationary programs without output dependencies. Matrix-multiply algorithms include Cannon~\citep{cannon1969cellular}, SUMMA~\citep{van1997summa}, and PUMMA~\citep{choi1994pumma}, COSMA~\citep{kwasniewski2019cosma}, Johnson~\citep{johnsonsalg}, and Solomonik~\citep{solomonik2011communication}. Early work by \citep{heathTriangular} explored distributed triangular solve, and \citep{kwasniewski2021costa} and \citep{ballard2009cholesky} investigated communication-optimal Cholesky. The distributed linear-algebra libraries ScaLAPACK~\citep{choi1992scalapack}, SLATE \citep{gates2019slate}, Elemental~\citep{poulson2013elemental} and DPLASMA \citep{bosilca2011flexible,bosilca2013parsec} implement variants of matrix multiply and solvers.


\section{Conclusion}
We present Cyclotron, a language and compiler that makes communication an expressible, first-class part of distributed tiled algorithms. By separating communication from computation and space--time mapping, Cyclotron lets programmers change how data moves without rewriting the computation itself, and makes it practical to explore a broad space of communication schedules. Across CPUs and GPUs, this flexibility allows Cyclotron to match and often outperform hand-optimized distributed linear algebra libraries, while showing that the best schedule depends on the machine, algorithm, and problem size. This work opens the door to deeper exploration of communication strategies for general networks of interacting processing elements, from multi-die GPU systems to wafer-scale engines and spatial dataflow accelerators. We believe this can enable software to better express and adapt to increasingly heterogeneous communication patterns as the hierarchies and interconnects of distributed computing platforms continue to diversify.

\section*{Acknowledgments}
The authors used generative AI tools to assist with code generation and debugging, as well as copyediting, wording, condensation, and proofreading. All technical ideas, experimental design, analysis, claims, and final decisions were produced and verified by the authors.

\bibliographystyle{ACM-Reference-Format}
\bibliography{references} 

\end{document}